\documentclass[final,3p,onecolumn,times,numbers]{elsarticle}

\usepackage{amssymb}
\usepackage{amsmath}
\usepackage{lipsum}
\usepackage{braket}

\begin{document}

\begin{frontmatter}

\title{Macroscopic "Lola/Mola" Cat State}

\author[first]{Harman Deep Kaur}
\author[second]{Mariagrazia Trapanese}
\author[third]{Kirill Zatrimaylov}
\affiliation[second]{organization={Ospedale Santa Chiara},
            addressline={Via Roma 67}, 
            city={Pisa},
            postcode={56126}, 
            country={Italy}}
\affiliation[third]{organization={Università degli Studi di Bergamo, Dipartimento di Ingegneria e Scienze Applicate},
            addressline={Viale Marconi 5}, 
            city={Dalmine (Bergamo)},
            postcode={24044}, 
            country={Italy}}

\begin{abstract}
We present the first--ever example of a macroscopic system in a quantum superposition. The system in question is a Siamese cat known as Lola; however, on a time scale of about 12 hours it oscillates into a different state that we refer to as "Mola". In the "Lola" state, the system is sweet and friendly and allows to cuddle itself, but in the "Mola" state, it is malevolent and witchy. When the probability of the system being in the "Mola" state is high, decoherence is strongly discouraged!

\end{abstract}

\end{frontmatter}

\section{Introduction}\label{introduction}
\begin{figure}
    \centering
    \includegraphics[width=0.5\linewidth]{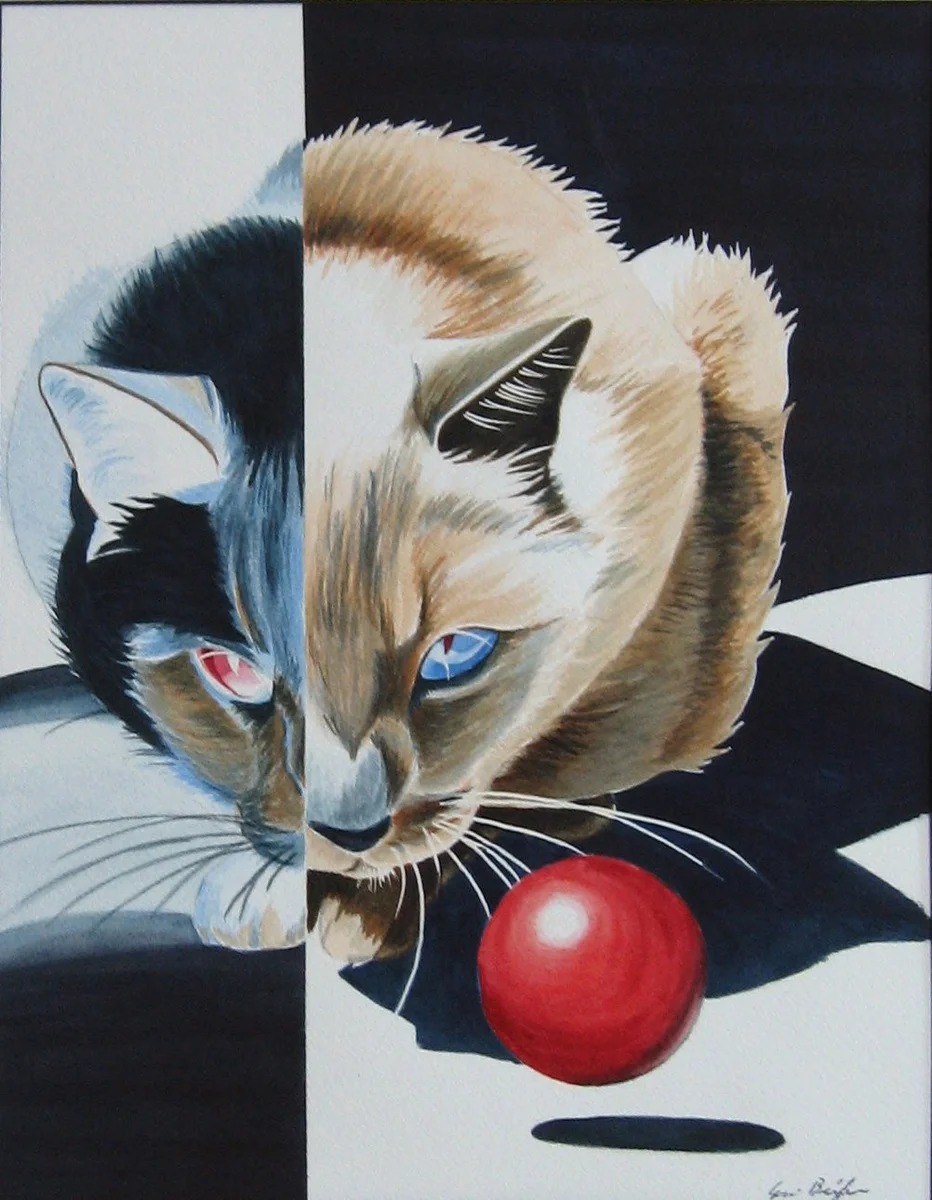}
    \caption{The "Lola/Mola" cat state (right side: Lola; left side: Mola).}
    \label{fig1}
\end{figure}

The issue of macroscopic superpositions of states has been a longstanding problem in the foundations of quantum mechanics. In particular, this was the subject of Erwin Schrödinger's famous thought experiment known as "Schrödinger's cat"~\cite{Schrodinger:1935zz}. The superposition property has been experimentally demonstrated for mesoscopic objects, including chiral molecules~\cite{Stickler:2021jyw}, nanodiamonds~\cite{Wood:2021icq}, and sapphire crystal resonators~\cite{Schrinski:2022ubf}.

Macroscopic superpositions are generally believed not to be observed in nature due to a process known as decoherence. namely thermodynamic interaction of the quantum system with the environment. This interaction results in a process known as einselection, making most quantum states macroscopically unstable and leaving out only specific states known as "pointer states"~\cite{Joos:1984uk,Zurek:2003zz}. However, it was shown in~\cite{Katsnelson:2017} that this is generally not true for dynamically fast environments. Computer simulations done in~\cite{Katsnelson:2018} also demonstrate that once the interaction between the system and the environment is turned off, the system regains its quantum properties.

In this paper, we present another counterexample to the decoherence paradigm, namely a macrosopic quantum two--state system that continuously oscillates between the two states. By an irony of fate, the system in question is an actual Siamese cat that lives with one of the authors of this paper, Mariagrazia Trapanese. The cat's name is Lola; however, experimental observations have established that she spontaneously changes into a completely different state that we called "Mola".

\section{Characteristic properties of the "Lola" and "Mola" states}\label{sec:LoMola}
\begin{figure}
    \includegraphics[width=0.5\linewidth]{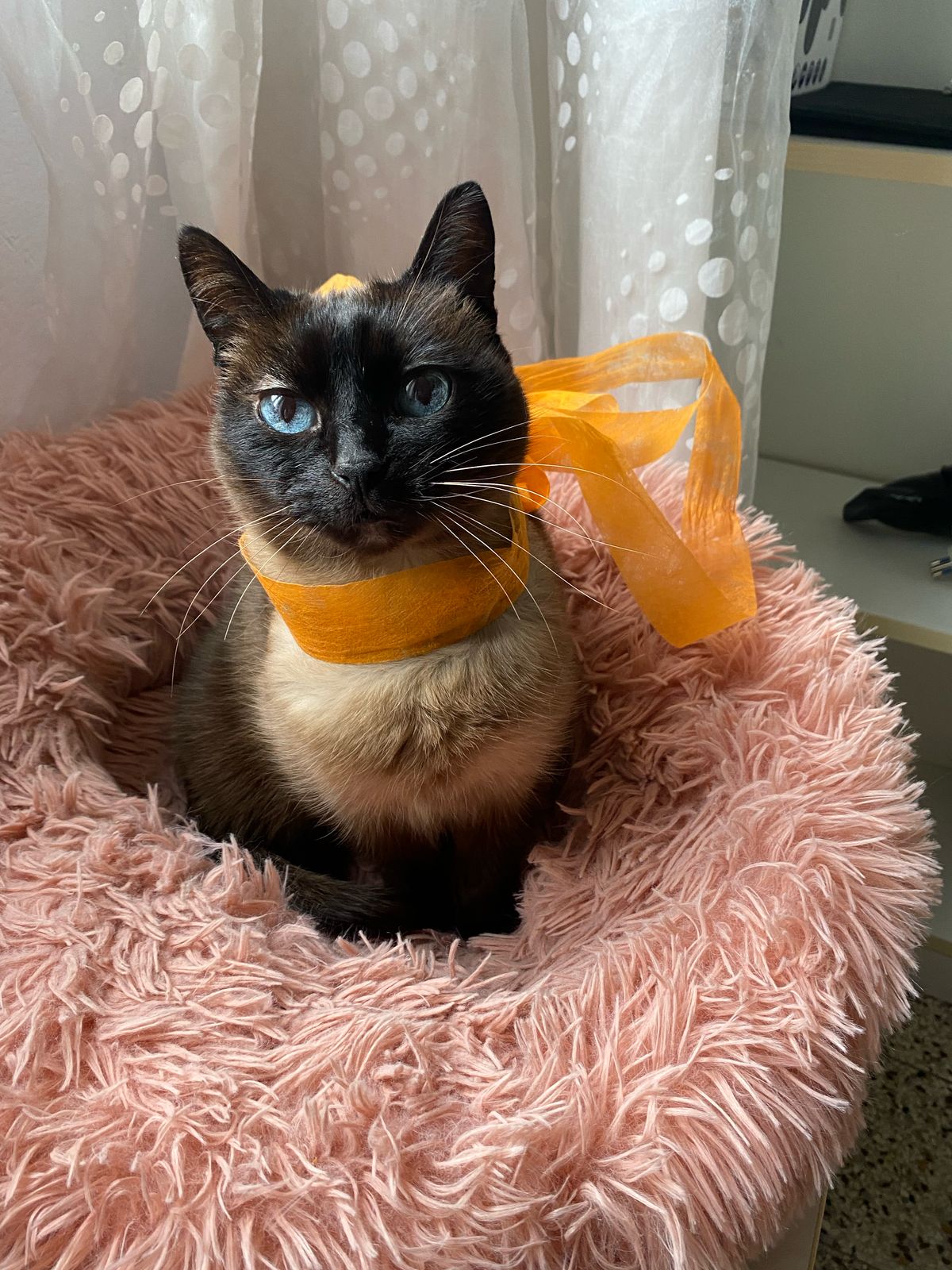}
    \includegraphics[width=0.5\linewidth]{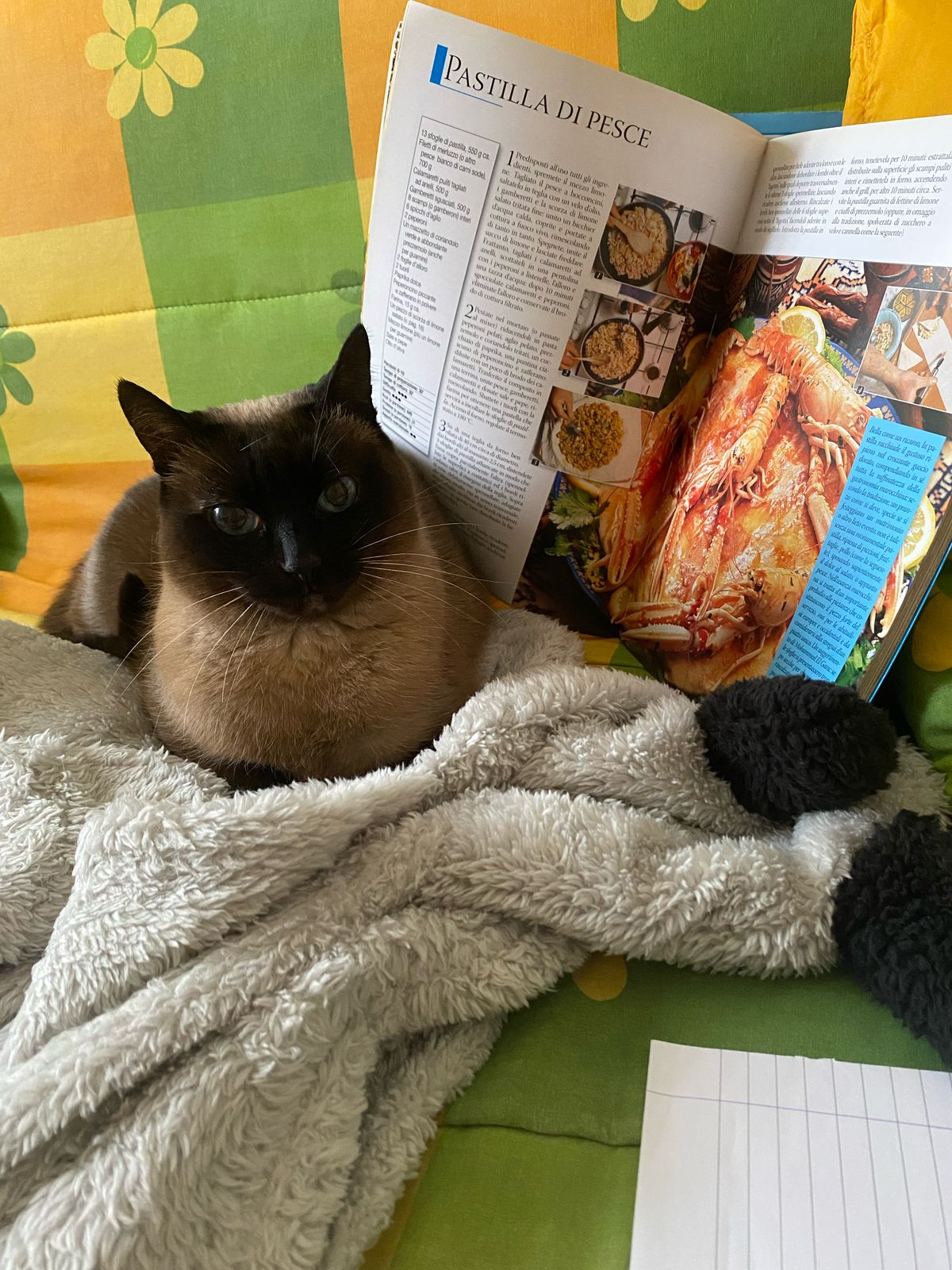}
    \includegraphics[width=0.4\linewidth]{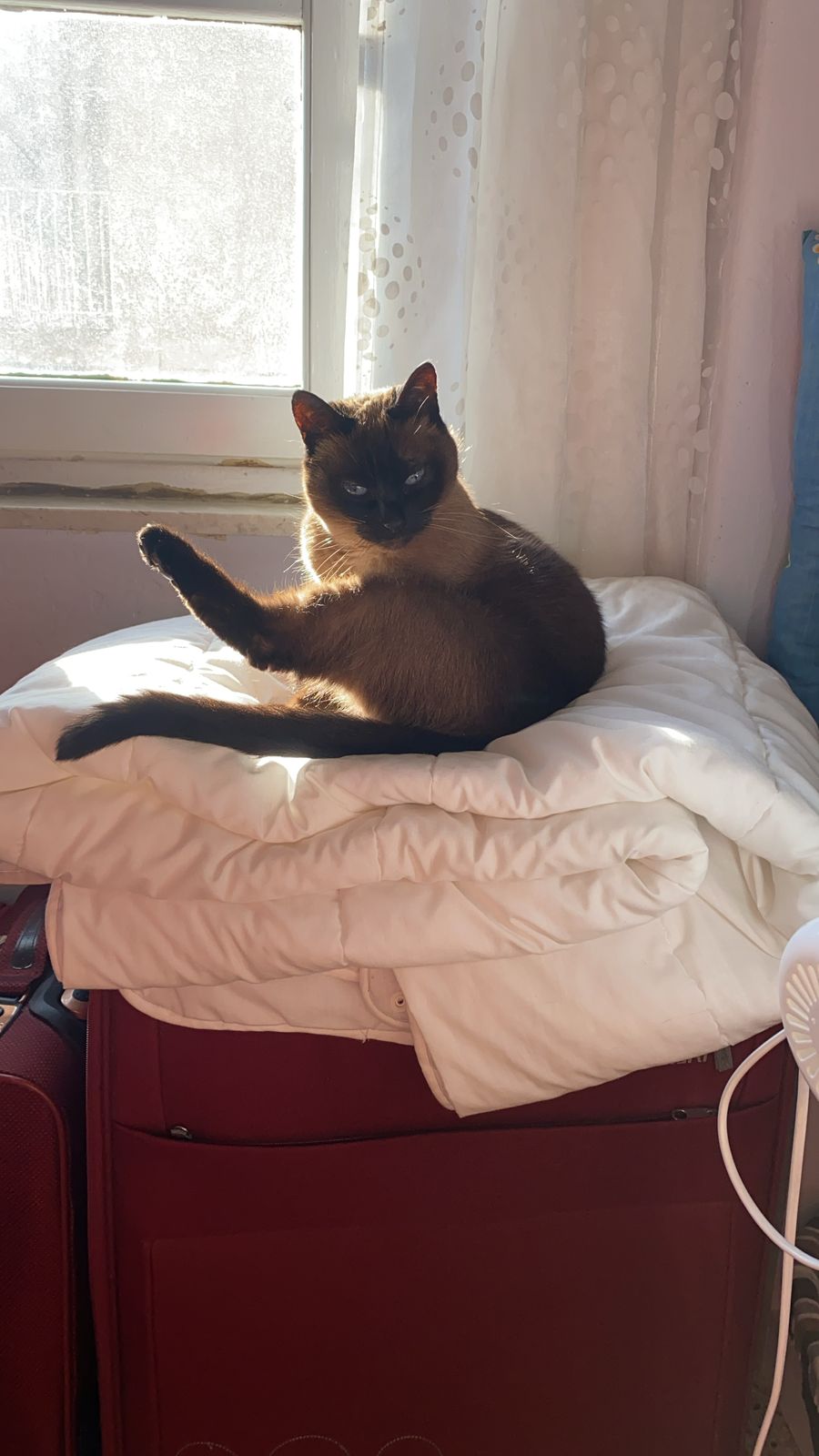}
    \includegraphics[width=0.5\linewidth]{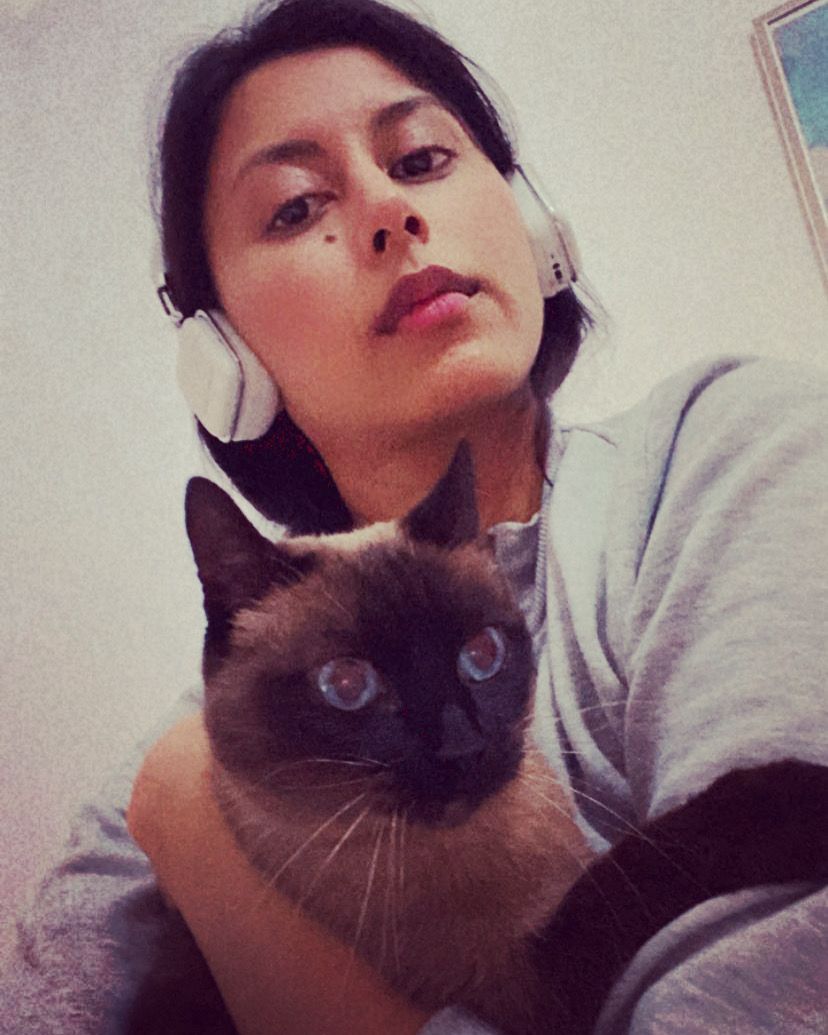}
    \caption{The system in the "Lola" state.}
    \label{Lola}
\end{figure}
\begin{figure}
    \includegraphics[width=0.5\linewidth]{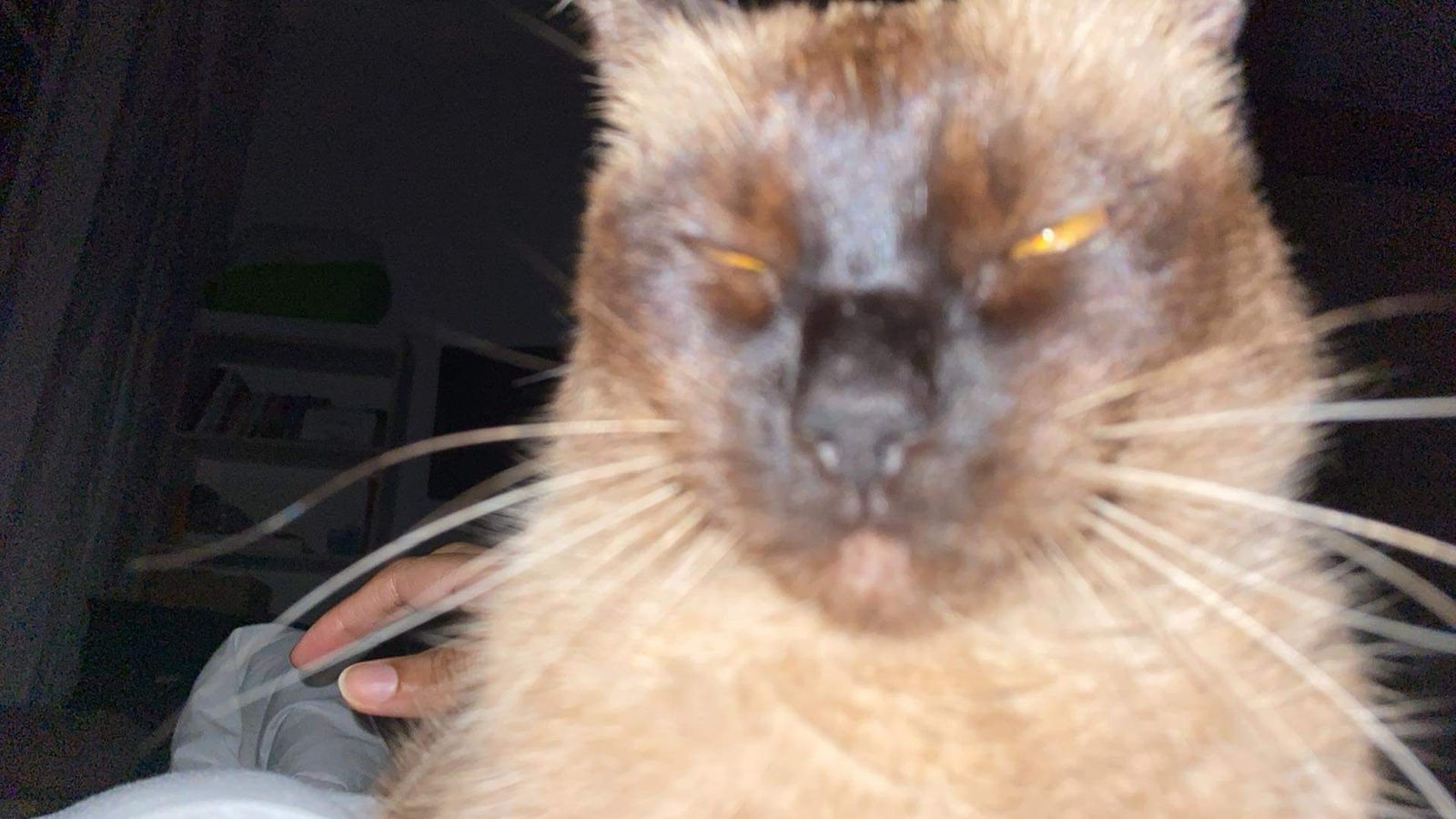}
    \includegraphics[width=0.5\linewidth]{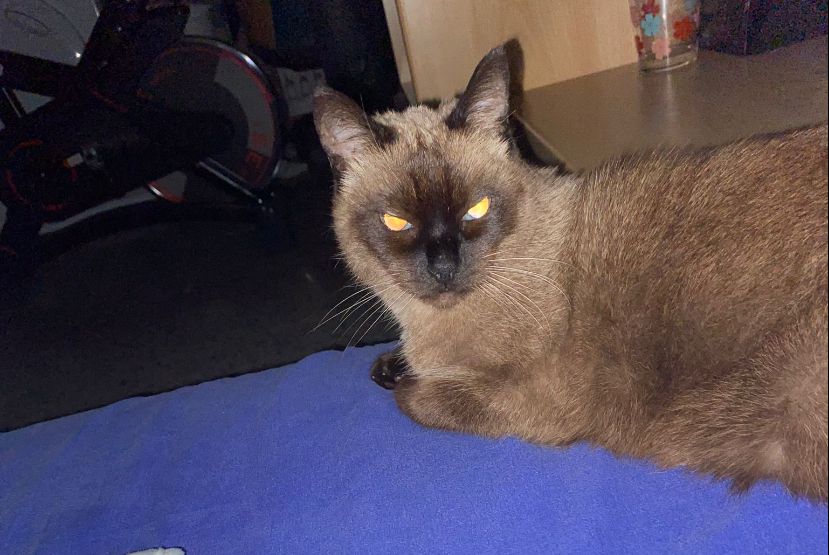}
    \includegraphics[width=0.6\linewidth]{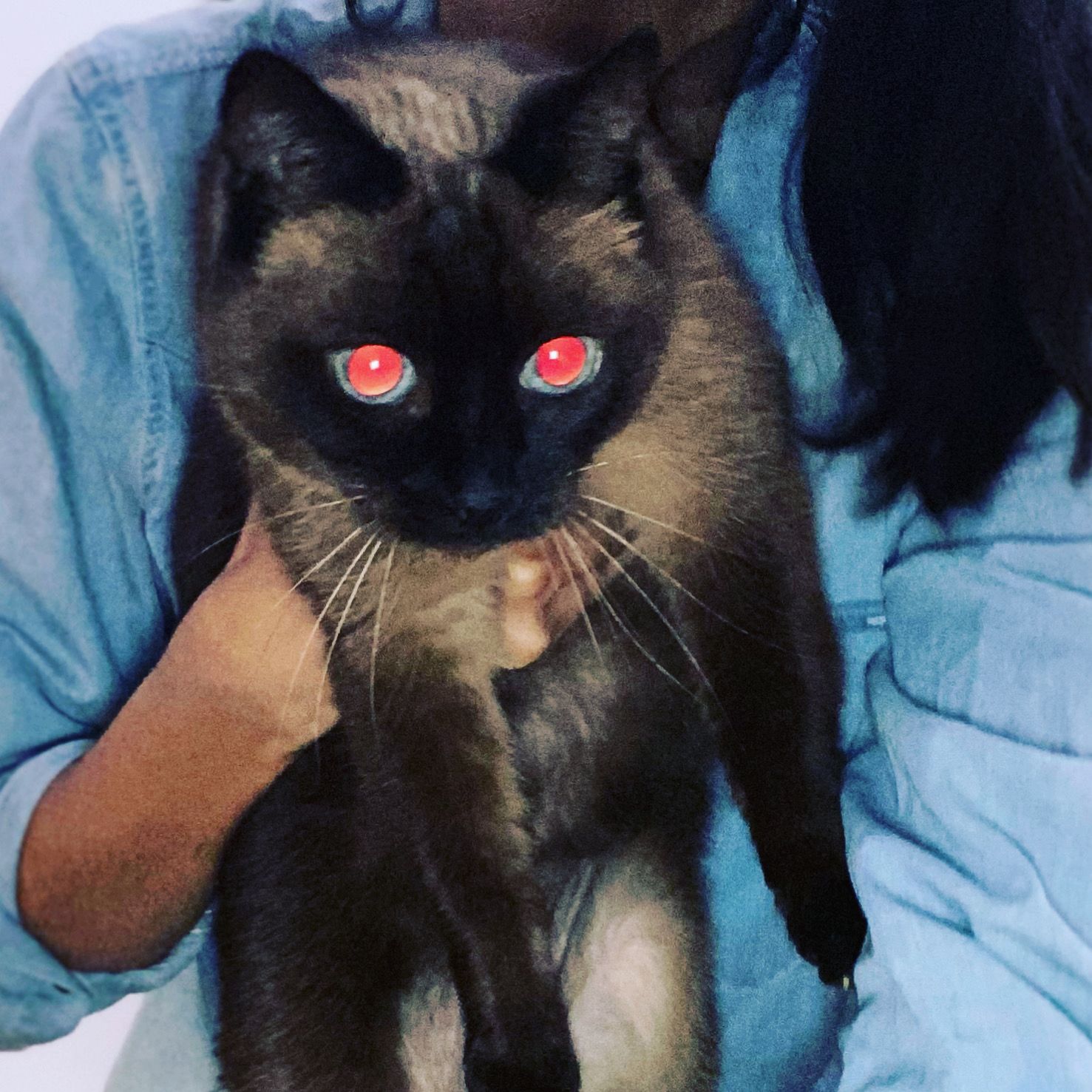}
    \includegraphics[width=0.5\linewidth]{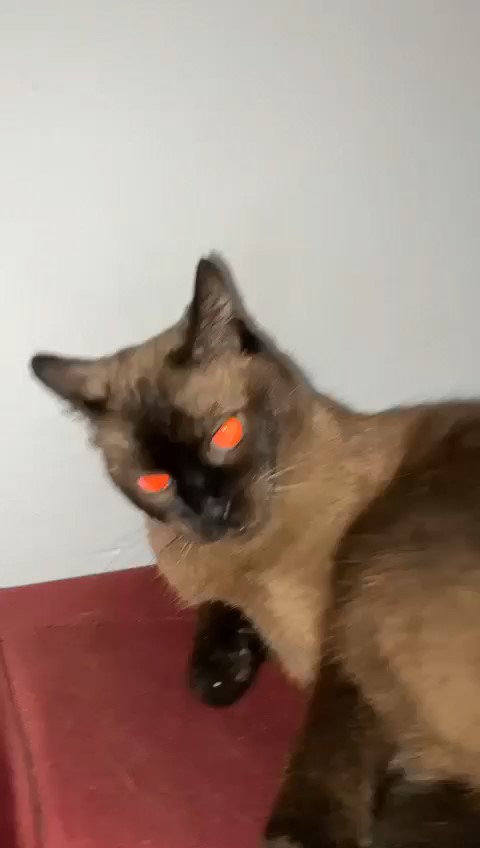}
    \caption{The system in the "Mola" state.}
    \label{Mola}
\end{figure}
The system is described by the wavefunction
\begin{equation}
\ket{\Psi(t)} \ = \ \cos\left(\frac{\pi}{2}\frac{t}{T}\right)\ket{LOLA}+\sin\left(\frac{\pi}{2}\frac{t}{T}\right)\ket{MOLA}
\end{equation}

The time scale $T$ has been estimated to be of order of about 12 hours: namely, at midday, the system is in the "Lola" state with about 100\% probability, while at midnight, it's in the "Mola" state with almost 100\% probability.

In the "Lola" state, the physical system is nice and friendly: it eats and plays, and allows to cuddle itself (see fig.~\ref{Lola}). Harman Deep Kaur, who performed experimental measurements on the system during daytime, reported having a extremely positive experience.

However, in the "Mola" state, the system displays completely different behavior: it is sinister, annoying, and cunning, and engages in black magic rituals (see fig.~\ref{Mola}). Harman Deep Kaur, who tried her best not to interact with the system during nighttime (and get some proper sleep), reported having a deeply unpleasant experience.

\section{Conclusions}\label{sec:conclusions}
In this paper, we have presented the first--ever example of a macroscopic system that demonstrates the property of quantum superposition and the ability to oscillate between two states.

During nighttime, when the Mola state is prevalent, it is also strongly discouraged to decohere the system.

\section{Acknowledgements}
We thank the system (in its "Lola" state) for its cooperativeness and willingness to help us study itself.

\section{Data availability}
No data was used for the research described in the article.

\end{document}